\documentclass[a4paper,11pt]{article}
\pdfoutput=1
\usepackage[T1]{fontenc}
\usepackage{pstricks}
\usepackage{enumerate}
\usepackage{color}
\usepackage{slashed}
\usepackage{graphicx,graphics}
\usepackage{verbatim}
\usepackage{amssymb}
\usepackage{amsthm}
\usepackage{array}
\usepackage{amsmath,amsfonts}
\usepackage{caption}
\usepackage[colorlinks=true,
            linkcolor=red,
            urlcolor=blue,
            citecolor=blue] {hyperref}
\usepackage[caption=false]{subfig}
\usepackage{cite}
\newcommand{\ba}{\begin{array}}
\newcommand{\ea}{\end{array}}
\newcommand{\bd}{\begin{displaymath}}
\newcommand{\ed}{\end{displaymath}}
\newcommand{\besub}{\begin{subequations}}
\newcommand{\eesub}{\end{subequations}}
\newcommand{\be}{\begin{equation}}
\newcommand{\ee}{\end{equation}}
\newcommand{\bea}{\begin{eqnarray}}
\newcommand{\eea}{\end{eqnarray}}

\def\l{\lambda}

\def\L{\Lambda}

\def\q2 {q^2}

\def\bt{\begin{table}}
\def\et{\end{table}}

\evensidemargin=\oddsidemargin
\catcode`@=11 
\def \gsim{\mathrel{\mathpalette\@versim>}}
\def \lsim{\mathrel{\mathpalette\@versim<}}
\def \@versim#1#2{\lower0.4ex\vbox{\baselineskip\z@skip\lineskip\z@skip
     \lineskiplimit\z@\ialign{$\m@th#1\hfil##\hfil$
     \crcr#2\crcr\sim\crcr}}}
\catcode`@=12 

\allowdisplaybreaks

\begin{document}
\begin{flushright}
	{\small HRI-RECAPP-2015-001}
\end{flushright}

\begin{center}
	{\Large \bf Dark matter, neutrino masses and high scale validity of an inert Higgs
		doublet model } \\
	\vspace*{1cm} {\sf Nabarun Chakrabarty
	$^{a,}$\footnote{nabarunc@hri.res.in}, ~Dilip Kumar Ghosh$^{b,}$\footnote{tpdkg@iacs.res.in}, ~Biswarup Mukhopadhyaya$^{a,}$\footnote{biswarup@hri.res.in}, ~Ipsita Saha$^{b,}$\footnote{tpis@iacs.res.in}} \\
	\vspace{10pt} {\small } {\em $^a$Regional Centre for Accelerator-based Particle Physics, Harish-Chandra Research Institute,  Chhatnag Road, Jhusi, Allahabad 211019, India \\
		$^b$Department of Theoretical Physics, 
		Indian Association for the Cultivation of Science, \\
		2A $\&$ 2B, Raja S.C. Mullick Road, Kolkata 700032, India}
	
	\normalsize
\end{center}

\begin{abstract}
	We consider a two-Higgs doublet scenario containing three $SU(2)_L$ singlet
	heavy neutrinos with Majorana masses. The second scalar doublet as well
	as the neutrinos are odd under a $Z_2$ symmetry. This scenario not only
	generates Majorana masses for the light neutrinos radiatively but also
	makes the lighter of the neutral $Z_2$-odd scalars an eligible dark matter 
	candidate, in addition to triggering leptogenesis at the scale of the heavy
	neutrino masses. Taking two representative values of this mass scale, 
	we identify the allowed regions of the parameter space of the model,
	which are consistent with all dark matter constraints. At the same time,
	the running of quartic couplings in the scalar potential to high scales
	is studied, thus subjecting the regions consistent with dark matter 
	constraints to further requirements of vacuum stability, perturbativity
	and unitarity. It is found that part of the parameter space is consistent
	with all of these requirements all the way up to the Planck scale,
	and also yields the correct signal strength in the diphoton channel for
	the scalar observed at the Large Hadron Collider. 
\end{abstract}	
\bigskip
 \section{Introduction}\label{intro}
The discovery of `a Higgs-like boson' \cite{Aad:2012tfa,Chatrchyan:2012ufa}, apparently completes the
Standard electroweak model (SM). However, it also compels us to mull over the
necessity of physics beyond the SM. To mention one motivation for this, a SM
Higgs of mass around 125 GeV, can break the absolute stability of vacuum at 
$10^{8-9}$ GeV, if the top quark mass $(M_t)$ and the strong coupling 
constant $(\alpha_s)$ are on the upper sides of their respective uncertainty bands.
A recent next-to-next-to-leading order (NNLO) study \cite{Degrassi:2012ry}
reveals that absolute stability up to the Planck scale requires
\begin{eqnarray}
M_h [\rm GeV] > 129.4 + 1.4(\frac{M_t[\rm GeV]-173.1}{0.7})-0.5(\frac{\alpha_s(M_Z)-0.1184}{0.0007})\pm1.0_{th} 
\end{eqnarray}
which, with all theoretical and experimental errors in $M_t$ and 
$\alpha_s$, implies that the absolute vacuum stability of SM is excluded at 95$\%$ 
CL for $M_h < 126$ GeV. To compensate the negative contribution 
from the fermion loops (mainly the top quark loop), which rapidly 
drives the quartic self-coupling of Higgs to below zero, we need
extra scalar loops that can ameliorate the problem. This provides one of
the important motivations to consider a theory with new scalars which
can make the vacuum stable up to high scales while having consistent 
results at low scales. 

At the same time it is worthwhile to remember two rather pressing issues 
which prompt one to look beyond the Standard Model (BSM). These are  
the non-zero mass and mixing of neutrinos and 
the likely existence of Weakly Interacting Massive Particle(WIMP),
contributing to Dark Matter (DM). While various Solar, atmospheric, reactor and accelerator-based
experiments provide strong evidence for non-zero masses of
neutrinos and mixing among their different flavors, 
several astrophysical and cosmological
observations suggest the existence of some exotic particles that constitute the DM 
content of the universe. Given all this, it is an attractive idea to look for a
theory that can simultaneously address all of the aforesaid problems in one framework. 
Here we consider one such scenario.

We investigate a model, first proposed in reference \cite{Ma:2006km}, that extends
the SM with an extra Higgs doublet and three right handed neutrinos
with a $Z_2$ symmetry, under which all SM particles are even while
this additional scalar doublet and the right-handed neutrinos are odd.
This symmetry prevents the additional doublet from having a vacuum 
expectation value(vev) thus vetoing the tree-level neutrino mass
generation. Moreover, there exists a stable scalar particle in the form of the
lightest neutral mass eigenstate of the additional doublet, which yields an eligible DM candidate.
The extra doublet is essentially an Inert Doublet.  
Although a lot of study has already taken
place on minimal Inert Doublet models 
\cite{Deshpande:1977rw,Barbieri:2006dq,LopezHonorez:2006gr,Cao:2007rm,Andreas:2009hj,Nezri:2009jd,Hambye:2009pw,Honorez:2010re,LopezHonorez:2010tb,
Dolle:2009ft,Miao:2010rg,Arhrib:2012ia,
Gustafsson:2012aj,Swiezewska:2012eh,
Goudelis:2013uca,Krawczyk:2013wya,Krawczyk:2013pea,Arhrib:2013ela,Arhrib:2014pva}, 
the extra appeal of this model lies in the
radiative generation of neutrino mass.
Though various aspects of this scenario have already been investigated
 \cite{Ma:2006fn,Krauss:2002px,Kubo:2006yx,Sierra:2008wj,Suematsu:2009ww,Suematsu:2010gv,Suematsu:2011va,Bouchand:2012dx,
Kashiwase:2012xd,Kashiwase:2013uy,Toma:2013zsa,Davoudiasl:2014pya}, the present study includes the 
following points which have not been emphasized before.
\begin{itemize}
	\item We investigate the vacuum stability 
	of this model at various scales, and identify the regions of its parameter space,
	which keeps the model valid all the way unto the Planck scale. The contribution of
	additional scalar fields as well as the right-handed neutrinos to the renormalisation group (RG)
	equations [given in Appendix \ref{ss:RGE}] has been taken into account here.
    Using these modified RG equations we evaluate the scalar quartic couplings
	at different scales. During the evolution of the quartic couplings we demand not only vacuum stability 
	but also perturbativity of the couplings as well as 
	unitarity of the $2\to2$ scattering matrix at each scale.
    
	\item  The heavy right-handed neutrinos introduce a new mass scale $(M)$ to the theory
	and the neutrino Yukawa couplings contribute to the RG evolution
	only at this mass scale. This brings out greater implications on the parameter space
	that distinguishes this model from minimal Inert doublet models. In this paper, we will show 
	the salient features of the model that emerges from the above fact. 
	
	\item  This model also contributes to the study of leptogenesis due to this heavy right-handed fermions. The values of the right-handed neutrino mass scale used in the high-scale
		analysis are taken to be commensurate with leptogenesis constraints. Thus the part
		of the model space consistent unto the Planck scale is also supportive of leptogenesis.
	
	\item We also examine the candidature of the lightest $Z_2$-odd particle as DM candidate,
	and identify the allowed values of the couplings yield the right relic abundance.
	We ensure that the DM candidate is consistent with the recent result of direct detection 
	experiments. We identify a substantial region of the parameter space, which simultaneously satisfy
	the vacuum stability, perturbativity and unitarity requirements and
	accommodate the DM candidate with the correct relic density.
	
	\item Finally, we 
	examine the 125 GeV scalar and make sure that the signal strengths in the observed channels (such as diphotons)
	are consistent with data from the Large Hadron Collider (LHC).  
	
\end{itemize}

We organize the paper as follows. In Section~\ref{model}, we briefly describe the 
model and its various features. In Section~\ref{constraints}, we explain all the
theoretical constraints and collider constraints that we 
use in the RG running of different quartic couplings.
Next, in Section~\ref{dmpart}, we discuss the DM aspects of this model. 
After explaining our analysis strategy in Section~\ref{analysis}, we present our results
related to high-scale validity in Section~\ref{results}. Finally 
in Section~\ref{conclusions}, we summarise our results.
\section{The Radiative Neutrino Mass Model with an Inert Doublet}\label{model}

In addition to the SM fields, the radiative neutrino mass model with an inert doublet \cite{Ma:2006km},
contains a Higgs doublet ($\Phi_2$) 
and three right handed (RH) neutrinos ($N_i$)   
with an unbroken $Z_2$ symmetry, under which the doublet and the right 
handed neutrinos are odd while all other SM particles are even. 
Being odd under the symmetry, $\Phi_2$ does
not acquire any vacuum expectation value (vev) and has no tree-level couplings to fermions. 

The relevant Yukawa and mass terms are 
\begin{eqnarray}
-{\cal L}_Y &=& y_{ij} \bar N_i {\tilde \Phi}^\dagger_2\ell_{j} + h.c 
+\frac{M_i}{2}\left({\bar N_i^c}N_i + h.c\right), (i,j=1-3)
\label{yukawa}
\end{eqnarray}

Here $\ell_i$ are the left-handed lepton doublets and $M_i$ are the Majorana mass term
for the heavy right-handed neutrinos $N_i$ \\
The scalar potential is
\begin{eqnarray}
V&=&\lambda_1({\Phi^\dagger_1}{\Phi_1})^2+\lambda_2({\Phi^\dagger_2}{\Phi_2})^2
+\lambda_3({\Phi^\dagger_1}{\Phi_1})({\Phi^\dagger_2}{\Phi_2}) 
+\lambda_4({\Phi^\dagger_2}{\Phi_1})({\Phi^\dagger_1}{\Phi_2}) \nonumber \\ 
&+&\Big[\frac{\lambda_5}{2}({\Phi^\dagger_1}{\Phi_2})^2 + {\rm h.c.}\Big]
+{m^2_{\Phi_1}}{\Phi^\dagger_1}{\Phi_1} + {m^2_{\Phi_2}}{\Phi^\dagger_2}{\Phi_2}
\label{potential}
\end{eqnarray}
where all parameters are real, and $\Phi_1$ is the SM Higgs doublet. 

The two scalar doublets can be written as 
\begin{eqnarray}
\Phi_1 = \left(\begin{array}{c}
G^+ \\ \frac{1}{\sqrt{2}}(v + h + iG) \end{array}\right) ~{\rm ~and,}~~ 
\Phi_2 = \left(\begin{array}{c}
H^+ \\ \frac{1}{\sqrt 2}(H + iA) \end{array}\right)   
\label{fields}
\end{eqnarray}
where, $v = 246$ GeV, is the electroweak vacuum expectation value (vev). One thus has five physical states $(h, H, A, H^\pm)$ and three
Goldstone bosons ($G, G^\pm$).
While $h$ corresponds to the physical SM-like Higgs field, the inert doublet components are one CP-even 
neutral scalar$(H)$, one CP-odd neutral scalar$(A)$ and a pair of charged scalars ($H^\pm$).
The physical masses are given by
\begin{eqnarray}
M^2_{H^{\pm}} &=& m^2_{\Phi_2} + \frac{1}{2}\lambda_3v^2 \nonumber \\
M^2_{H} &=& m^2_{\Phi_2} + \lambda_Lv^2 \nonumber \\
M^2_{A} &=& m^2_{\Phi_2} + \lambda_Av^2 
\label{eq:scamass}
\end{eqnarray}
where $\lambda_{L/A} = \frac{1}{2}(\lambda_3 + \lambda_4 \pm \lambda_5)$. The value of $\lambda_1$ is
determined using $M_h = 125$ GeV.\\
Majorana masses for the light neutrinos are generated radiatively through one-loop exchange of the $Z_2$-odd 
neutral scalars. The general expression for the loop-induced contributions to the light neutrino mass matrix \cite{Ma:2006km} is
\begin{equation}
{\cal M}^\nu_{ij}=\sum_{k=1}^3\frac{y_{ik} y_{jk} M_k}{16\pi^2}
\left[\frac{M^2_{H}}{M^2_{H}-M^2_k}
\ln\frac{M^2_{H}}{M_k^2}-\frac{M^2_{A}}{M^2_{A}-M^2_k}
\ln\frac{M^2_{A}}{M_k^2}\right]
\label{nmass}
\end{equation}
Thus, the neutrino masses and mixing are determined by the inert scalar masses and the right-handed
neutrino masses $M_i$. These masses represent the scale of lepton number violation and hence that 
of leptogenesis \cite{Fukugita:1986hr,Pilaftsis:1997jf} in this model. Some studies have already been done in this context \cite{Hambye:2009pw,Suematsu:2011va,Kashiwase:2012xd,Kashiwase:2013uy,Racker:2013lua}.  Our choice of the right-handed neutrino (Majorana)
	mass scales made our study automatically compatible with leptogenesis. To satisfy the necessary
	constraints in the low DM mass region where $M_{DM}<M_W$, one must take the lightest of the Majorana
	masses to be $M \geq 110 ~\rm TeV$, where as in the high DM mass region ($M_{DM}>500 ~\rm GeV$)  the bound is only 1 TeV \cite{Hambye:2009pw}. 
	Hence, to be consistent in both cases, we use two values of M, 
(a) $M = 110 ~\rm TeV$ and (b) $M = 10^{9} ~\rm TeV$. While, in one hand, choice of (a) is motivated by the 
idea of having the lowest possible leptogenesis scale, we choose to work with (b) which
have interesting consequences on the RG runnning. We will show in later sections, how
the mass scale (b) of $M$ affects the stability of the vacuum mainly in the
high DM mass region and eventually explain the physical reasons behind it.   

Along with the above restrictions we also demand $M_\nu\sim{\cal O}(0.1~\rm eV)$ to be consistent
	with neutrino oscillation data for some fixed $M$ and other exotic scalar masses. However, for simplicity, we consider only one diagonal Yukawa coupling $(y_\nu)$ and do not look into the hierarchical details of the Yukawa matrix. At this point, it is to be noted 
that a more rigorous study with the intricate flavor 
structure of the neutrino Yukawa matrix can highlight the region of the
model space that fits the observed pattern of neutrino mixing \cite{Davoudiasl:2014pya}. However, 
we would like to emphasize that the 
broad conclusions on the high-scale validity of this scenario vis-a-vis the DM
constraints remain unchanged.
Finally, we should mention that the lighter state between H and A is the DM candidate. We present our illustrative results for cases where H plays this role.

\section{Constraints from perturbativity, unitarity, vacuum stability and collider data}\label{constraints}
In this section, we briefly describe the constraints that are imposed on the model parameters
and how exactly they shape the results so obtained.

\subsection{Vacuum stability}
The scalar potential is considered bounded from below, 
if it does not turn negative for large field values along all possible field directions. 
In this model, stability of the electroweak vacuum is ensured up to some specified energy 
scale if the following conditions are satisfied for all scales Q up to that scale: 
\besub 
\bea
\label{e:vsc1}
\rm{vsc1}&:&~~~\lambda_{1}(Q) > 0 \\
\label{e:vsc2}
\rm{vsc2}&:&~~~\lambda_{2}(Q) > 0 \\
\label{e:vsc3}
\rm{vsc3}&:&~~~\lambda_{3}(Q) + \sqrt{\lambda_{1}(Q) \lambda_{2}(Q)} > 0 \\
\label{e:vsc4}
\rm{vsc4}&:&~~~\lambda_{3}(Q) + \lambda_{4}(Q) - |\lambda_{5}(Q)| + \sqrt{\lambda_{1}(Q) \lambda_{2}(Q)} > 0 
\eea
\label{eq:vsc}
\eesub
Such conditions have been elaborately discussed in literature 
\cite{Sher:1988mj,Nie:1998yn,Ferreira:2004yd,Branco:2011iw}. 
One should make a note that these conditions ensure \emph{absolute} 
stability of the electroweak vacuum. For metastability, the conditions are somewhat less stringent.

\subsection{Perturbativity}
For the scalar quartic coupling $\lambda_i(i=1-5)$, the criterion for perturbativity is,
\bea
\label{pert:stable}
\l_i(Q) < 4\pi 
\eea
The corresponding constraints for the Yukawa and gauge interactions are,
\bea
y_i(Q), ~g_i(Q) < \sqrt{4\pi}
\eea
where, Q represents the energy scale at which they are being computed. 
We demand that the criteria in Eq.~\ref{pert:stable} be obeyed at all energy scales 
up to the cut-off of this model.

\subsection{Unitarity} 
A further set of conditions come on demanding unitarity of the scattering matrix 
comprising all $2\to2$ channels involving, by the optical theorem \cite{Ginzburg:2005dt,Gorczyca:2011he,Bhattacharyya:2013rya,Chakrabarty:2014aya}. 
In our context, this translates into the condition that each distinct eigenvalue of the aforementioned amplitude matrix be bounded above at 8$\pi$ (after factoring out $1\over16\pi$ from the matrix). These eigenvalues are:

\begin{eqnarray}
a_\pm &=& \frac{3}{2}(\lambda_1 + \lambda_2)\pm\sqrt{\frac{9}{2}(\lambda_1 - \lambda_2)^2 + (2\lambda_3+\lambda_4)^2} \nonumber \\
b_\pm &=& \frac{1}{2}(\lambda_1+\lambda_2)\pm\sqrt{\frac{1}{4}(\lambda_1-\lambda_2)^2 + \lambda^2_5}, \nonumber \\
c_\pm &=& d_\pm = \frac{1}{2}(\lambda_1+\lambda_2)\pm\sqrt{\frac{1}{4}(\lambda_1-\lambda_2)^2 + \lambda^2_5}, \nonumber \\
e_1 &=& (\lambda_3 + 2\lambda_4 - 3\lambda_5), \nonumber \\
e_2 &=& (\lambda_3 - \lambda_5), \nonumber \\
f_1 &=& f_2 = (\lambda_3 + \lambda_4), \nonumber \\
f_+ &=& (\lambda_3+2\lambda_4+3\lambda_5), \nonumber \\
f_- &=& (\lambda_3 + \lambda_5). 
\end{eqnarray}

\subsection{Collider data}
In addition to the theoretical constraints discussed above, important bounds on scalar mass parameters come from collider data. 

\begin{itemize}
  \item  In order to identify $h$ with
the SM-like Higgs as observed by the ATLAS and CMS collaborations, 
one requires $M_h\simeq125$ GeV.

  \item 
To be consistent with the LEP bounds, one must have
\begin{eqnarray}
M_{H} + M_{A} > M_Z\,,\\
M_{H^{\pm}} + M_{A/H} > M_W\,. \nonumber
\end{eqnarray}    
Moreover, for neutralino as in the supersymmetric context, LEP-II searches limit pseudo-scalar mass $(M_A)$ to 100 GeV
when $M_{H} < M_{A}$ \cite{Cao:2007rm,Lundstrom:2008ai}. Similarly, chargino search data, properly extrapolated, imply $M_{H^\pm}>70$ GeV 
\cite{Pierce:2007ut}.
  
  \item Though all the tree-level couplings of $h$ are identical to those of the SM Higgs,
  the charged scalar $H^{\pm}$ potentially modifies the loop induced couplings
  $h \gamma \gamma$ and $h Z \gamma$ via loop contributions \cite{Shifman:1979eb,Ellis:1975ap,Cahn:1978nz,Bergstrom:1985hp,Djouadi:2005gj}.  
  We theoretically compute the signal strength $\mu_{\gamma \gamma}$ for $h$ 
  decaying to the diphoton channel as the ratio of the decay width in the {\it `model'} to that in the SM. 
Its experimental value reported by the ATLAS and CMS collaborations stand at $1.17 \pm 0.27$ and 
$1.13 \pm 0.24$ respectively \cite{Aad:2014eha,Khachatryan:2014ira}. 
 Demanding the signal strength to be within $2\sigma$ limits of 
the experimentally quoted values puts further constraints on the model. We use
the limit on $\mu_{\gamma \gamma}$ weighted as below:
\bea
{1\over {\sigma^2}} = ({1\over {\sigma^2}})_{ATLAS}+({1\over {\sigma^2}})_{CMS}
\label{mu_gamma}\\
{\mu_{\gamma\gamma}\over {\sigma^2}} = ({\mu_{\gamma\gamma}\over {\sigma^2}})_{ATLAS}+({\mu_{\gamma\gamma}\over {\sigma^2}})_{CMS}
\label{mu_gamma}
\eea  
where, the numerators on the right hand side denotes the central values of the respective 
experimental results and $\sigma$ are the corresponding uncertainties.
\end{itemize}

\section{Dark Matter Issues}\label{dmpart}

As stated earlier, we identify $H$ as the DM candidate. For complimentary choice, namely,
$A$ with the same mass as the DM candidate, we have checked that the contribution to the
relic density is of similar magnitude.
The relevant DM constraints to be considered are as follows:  
\begin{itemize}
\item According to recent PLANCK experiment\cite{Ade:2013zuv}
the observed cold DM relic density is
\begin{eqnarray}
\Omega_{\rm DM}h^2 = 0.1199\pm0.0027
\label{planck_data}
\end{eqnarray}    
We restrict our result up to $3\sigma$ deviation from the central value. 
\item Strong constraints come from direct DM searches. 
Recently, XENON100\cite{Aprile:2012nq} and LUX\cite{Akerib:2013tjd}
experiments have put upper-bound on the DM-nucleon scattering cross-section for a 
wide range of the DM mass. In our case, the direct detection cross-section 
strategy is based on t-channel Higgs mediation. We choose to work in the 
region of the parameter space allowed by the LUX limit.
\item For $M_H \leq M_h/2$, 
the decay mode of Higgs to two DM particle $(h \to H H)$ will
presumably contribute to the invisible decay width of the Higgs boson. 
We take into account the current constraint on the Higgs invisible 
branching ratio $(< 20\%)$ from model independent Higgs precision analysis 
\cite{Banerjee:2012xc,Cheung:2014noa}. 
\item Our parameter scan has included situations where the $Z_2$-odd scalar and the pseudo-scalar
are close by in mass and with co-annihilation yield correct relic abundance within the $\pm3\sigma$
of the observed values when the DM mass is high ($> 500 \rm~ GeV$). For DM mass in between
100 GeV and 500 GeV,  
there is {\it prima facie} the possibility of co-annihilation between the
spinless DM candidate and a right-handed neutrino which can give rise to correct relic abundance, 
as discussed in Ref.~\cite{Klasen:2013jpa}. 
In our case, however, such a possibility is ruled out by the additional
requirement of leptogenesis, which in turn puts a lower bound on the mass
range for the right-handed neutrino(s).
\end{itemize}
Hence, only the following two cases survive all the above constraints. 

\subsection{Case-A: Low mass DM ($50 \rm~GeV < M_H < 90 \rm~GeV$)}\label{casea}

In this mass region, the dominant annihilation channel for H goes to 
the DM self-annihilation processes  
through $h$ mediation.
This keeps the relic density at the right level. 
For both positive and negative values of DM-DM-Higgs couplings, the 
relic density remains within the allowed range as long as the
$M_H < 90$ GeV. The sub-dominant contribution to the 
relic density comes from the t-channel processes to vector
boson final states mediated by $A$ and $H^\pm$. A detailed
discussion in this regard on a similar model can be found,
for example in \cite{Hambye:2009pw,LopezHonorez:2010tb,Arhrib:2013ela}.
However, the coupling $\lambda_2$ 
has no effect in the relic density calculation. In the next section, we will discuss
this results elaborately. One more notable point is that, 
annihilation processes that mediated by the heavy right-handed neutrinos
give negligible contribution (less than $1\%$) to the relic density
calculation. These processes are suppressed by the heavy mediator mass.

\subsection{Case-B: High mass DM ($M_H > 500 \rm~ GeV$)}\label{caseb}

The interesting feature of this region is that, the correct
relic abundance can be achieved if and only if $H$, $A$ and
$H^\pm$ are almost degenerate, at most have a mass difference
of the order of 10 GeV. This is mainly because at this high mass, 
the annihilation channels with vector boson final states can have direct 
quartic couplings $(HHVV,V=W^\pm,Z)$ or can be mediated by any of the three scalars through
$t/u$ channels. These diagrams yield too large an annihilation cross-section to match with
the proper relic density. However, cancellation between direct quartic coupling
diagrams and $t/u$ channels diagram occurs for mass-degenerate
$H$, $A$ and $H^\pm$, which in turn brings down
the annihilation cross-section to the desired range. 

\section{Analysis strategy}\label{analysis}

The aim of this study is to probe the parameter space of an inert 
doublet model (IDM) augmented with heavy RH
neutrinos compatible with various theoretical and experimental constraints
elaborated in the previous sections. We carry out our investigation in the two 
separate mass regions. In each region, we scan over the relevant
parameters, namely, the masses $M_H$, $M_A$ and $M_H^{\pm}$, and the coupling $\l_L$.
With $M_h$ fixed at 125 GeV, $\l_2$, $\l_L$, $M_H$, $M_A$ and $M_H^{\pm}$ 
fix all the remaining quartic interactions.  
 At this point it is to be mentioned 
that the parameter $\lambda_2$ can not be constrained by any physical observable at least
at tree level. However, $\lambda_2$ is bounded by the stability and perturbativity condition 
and we have explicitly checked that $0 < \lambda_2 < 0.36$ to satisfy the theoretical 
constraints. Rather than scanning over $\lambda_2$, we have
demonstrated our results with two benchmark values (0.1 and 0.001) for it, both
of which are well within the above limit. 

These quartic couplings are then used as the electroweak boundary conditions at  
$Q = M_t$ and their RG evolution to high scales is studied. Here $M_t$ denotes
the top quark pole mass. The reader is reminded that the effect of the 
RH neutrinos is 
turned on at a scale $Q = M$. Thus, for $M_t \le Q \le M$, we do not include the RH
neutrino contributions to the one-loop RG equations. We include such conditions for $Q \ge M$ and use the
RG equations listed in the appendix.
The scale up to which the scenario remains consistent is denoted by 
$\L_{UV}$. For a generic parameter point {$\l_i(Q = M_t)$}, we check the 
aforementioned theoretical
constraints at all intermediate scales up to $\L_{UV}$. If 
the constraints are all satisfied, we tag {$\l_i(Q = M_t)$}
as an \emph{allowed} point. This marks out an allowed region in the parameter space
defined at the electroweak scale. Moreover, the effects of constraints 
stemming from the DM observables and collider searches are 
examined independently in this region. The finally allowed parameter regions are
thus identified and presented for benchmark values of $\l_2$ and $M$. 
We use the publicly available package micrOMEGAs-v3.6.9.2 \cite{Belanger:2013oya} for DM analysis.
In all the cases, we also make sure that our allowed parameter space does not violate the oblique T-parameter constraints \cite{He:2001tp,Grimus:2007if}.

Amongst the SM fermions, only the top quark plays the dominant role in the evolution of the couplings.
The boundary condition for its Yukawa interaction at the electroweak scale is fixed by $y_t(M_t) = \frac{\sqrt2 M_t}{v} (1 - \frac{4}{3 \pi} \alpha_s(M_t))$. We have used $M_t$ = 173.39 GeV throughout our analysis.

\section{Numerical results}\label{results}

\subsection{Low mass DM region}

We perform a detailed parameter space scan where $M_H < 100$ GeV. In this scan, we impose the
LEP bounds as discussed in Sec 3.4.
\bea
\l_L: [-1.0,1.0] \\
M_H:  [50.0~\rm GeV, 90.~\rm  GeV] \\
M_A : [100.~\rm  GeV, 500.~\rm  GeV]\\
M_{H^+} : [100.~\rm  GeV, 500.~\rm  GeV]
\eea 

\begin{figure}[h!]
\begin{center}
    \subfloat[ \label{sf:t2m0l}]{%
      \includegraphics[scale=0.5]{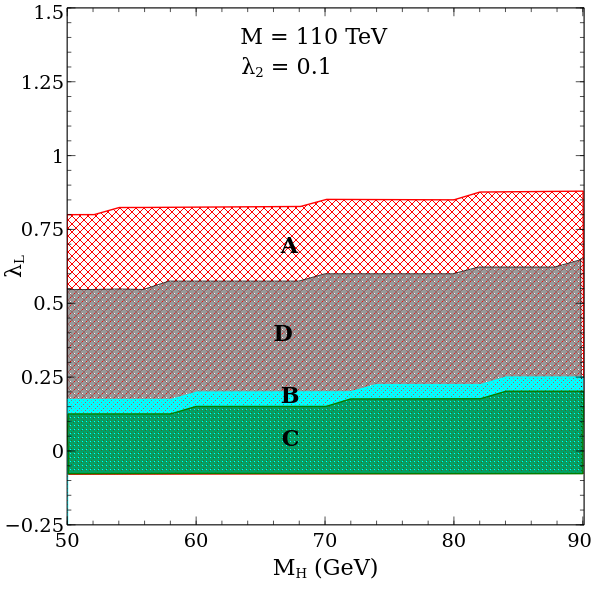}
    }~~~~
    \subfloat[\label{sf:t2m0LQT}]{%
      \includegraphics[scale=0.5]{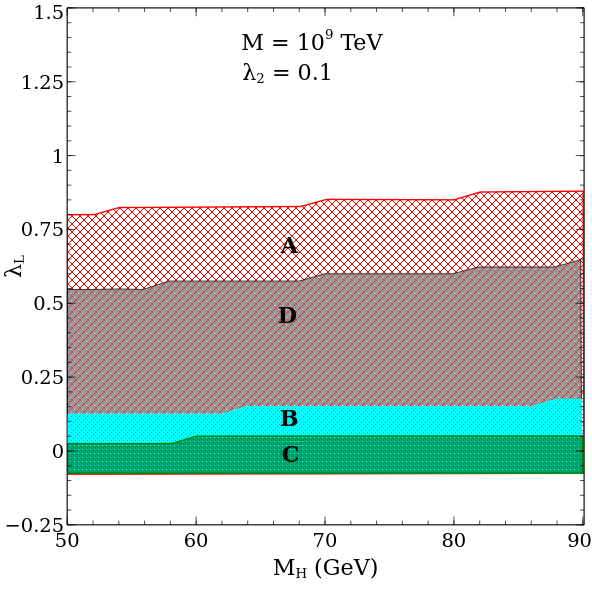}~~~~
    }
    \vspace*{0.2cm}
        \subfloat[ \label{sf:t2m0l}]{%
      \includegraphics[scale=0.5]{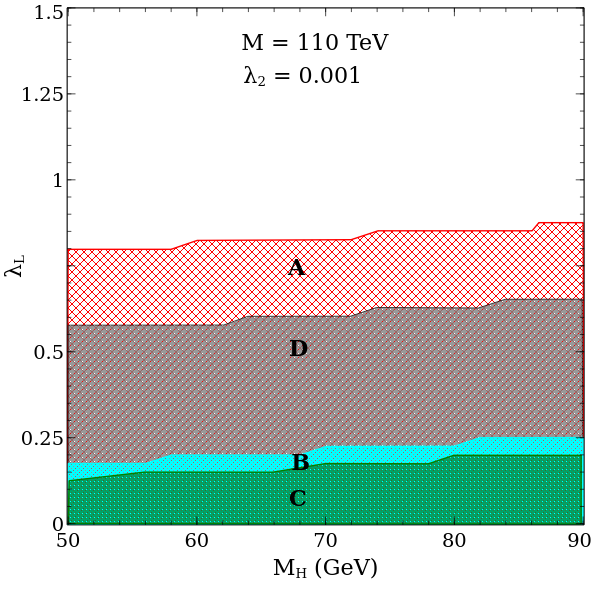}
    }~~~~
    \subfloat[\label{sf:t2m0LQT}]{%
      \includegraphics[scale=0.5]{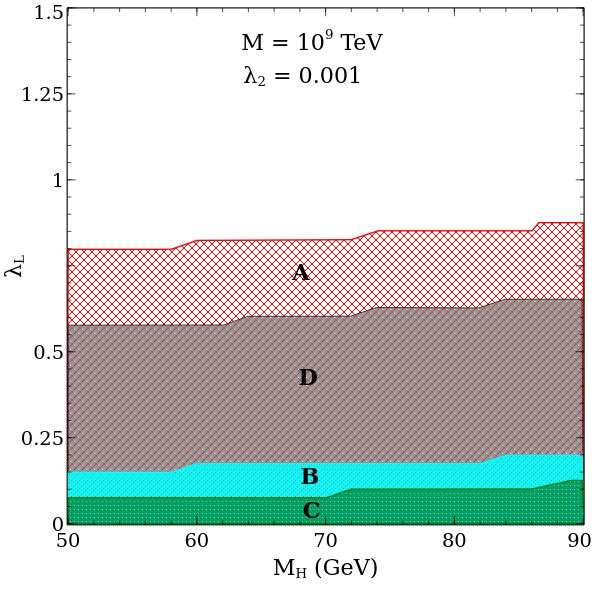}~~~~
    }
\end{center}
\caption{Regions compatible with the theoretical constraints in the $M_H$-$\lambda_L$ plane
	for $M = 110$ TeV (left panel) and $10^{9}$ TeV (right panel)
	with three different choices of $\Lambda_{UV}$ and two values of $\lambda_2$ (upper and lower panel) . 
	The regions denoted by A (red), B (cyan) and 
	C (green) obey these constraints up to $\Lambda_{UV}$ = $10^{6}$, $10^{16}$ and $10^{19}$ GeV, respectively. 
	The grey region denoted by D keeps the Higgs to diphoton signal strength within 2$\sigma$ limits of the current data.}
\label{f:lLMH}
\end{figure}

We solve the RG equations for two values of $M$, $110$
TeV and $10^{9}$ TeV respectively. 
We then show the allowed parameter space in the $\l_L-M_H$ plane
for different choices of 
$\L_{UV}$ in Fig \ref{f:lLMH} and \ref{f:lLMH_DM}. The regions denoted by A (red), B (cyan) and 
	C (green) denote $\L_{UV}$ = $10^{6}$, $10^{16}$ and $10^{19}$ GeV respectively. 
	We overlay the region allowed by the Higgs to diphoton signal strength within 2$\sigma$ limits of the current data 
	as grey region named D. 
As mentioned earlier, the full analysis is done  
for two values of $\lambda_2$ at the electroweak scale (0.1 and 0.001).

Let us briefly summarize the features that emerges from the figures and their possible explanations. 
\begin{itemize}
	\item From Fig.\ref{f:lLMH}, it can be seen that an Inert doublet model with 
	right handed neutrinos posses stable vacuum even up to the Planck scale, but
	the higher the cut-off scale $\L_{UV}$, the less amount of parameter space becomes allowed. To understand
	this, one can recollect that the top quark Yukawa coupling in the SM is responsible for the downward evolution of the 
	scalar self-coupling, which poses a threat to the vacuum stability. The presence of the 
	additional scalar quartic couplings ($\l's$) in a model like this offsets such an effect; however, the boost
	thus provided to these couplings tend to violate the perturbativity and 
	unitarity condition. This necessitates a tightrope walking, and the scale up to
	which it is possible is $\L_{UV}$. Hence, it is natural that for higher $\L_{UV}$, fewer combinations of parameters will achieve this fine balance, and consequently the allowed region shrinks. 
	\item Fig.\ref{f:lLMH} also shows that $\l_L$ is bounded on both sides and the limits stay almost same for different
	right handed neutrino mass scale$(M)$. This is of no surprise
	and can be easily understood. Since $\l_3 = \frac{2}{v^2} (M_{H^\pm}^2-M_H^2+\l_L v^2)$, the upper bound on $\l_L$ is imposed by the requirement of perturbative unitarity. This is because a higher positive value of $\l_L$
	makes $\l_3$ large at the electroweak scale which violates perturbative unitarity
	in the course of its evolution under RG. On the other hand, a large negative value
	of $\l_L$ induces a large negative value to $\l_3$. As a consequence, the 
	vacuum stability condition vsc4 of Eq.\ref{e:vsc4} is violated even near the electroweak(EW) scale. 
	This puts a lower limit on $\l_L$ independent of $M$ and $\L_{UV}$, as
	evident from Fig.\ref{f:lLMH}.  However, it must be noted that the lower limit
		of $\l_L$ is not independent of $\lambda_2$, which is another consequence of Eq.\ref{e:vsc4} that  
		requires the condition $\lambda_3 + \lambda_4 - |\lambda_5| > -\sqrt{\l_1\l_2}$  to be
		satisfied. With the decrease in the value of $\l_2$ the lower limit of the combination $\lambda_3 + \lambda_4 - |\lambda_5|$ enhances. This combination can be approximated to $\l_L$ in the perturbative regime and hence lower values of
		$\l_2$ increases the lower bound of $\l_L$, as can be seen from the upper and lower panel of Fig.\ref{f:lLMH}
		that corresponds to $\l_2=0.1$ and $\l_2=0.001$ respectively.
	\item To check the compatibility of the DM constraints with the theoretical ones,
	we look for the region allowed by the 3$\sigma$ limits on $\Omega_{\rm DM}h^2$ from PLANCK data
	and 90$\%$ CL limit on the spin-independent DM-nucleon scattering cross-section from
	LUX data. In Fig.~\ref{f:lLMH_DM}, we show the parameter space allowed by the entire set of DM constraints.
	An inspection of Fig.~\ref{f:lLMH} and Fig.~\ref{f:lLMH_DM} shows that almost the full parameter space 
	allowed by the DM constraints lies in the region which is also favoured by the vacuum stability condition all
	the way up to the Planck scale. However, the region corresponding to
	$\l_L \le -0.1$ (shaded region in Fig.~\ref{f:lLMH_DM}) does not lead to the stable vacuum. 
	\begin{figure}[htb!]
		\centering
		\includegraphics[trim = 50 155 70 150,clip,width=10cm,height=10cm]{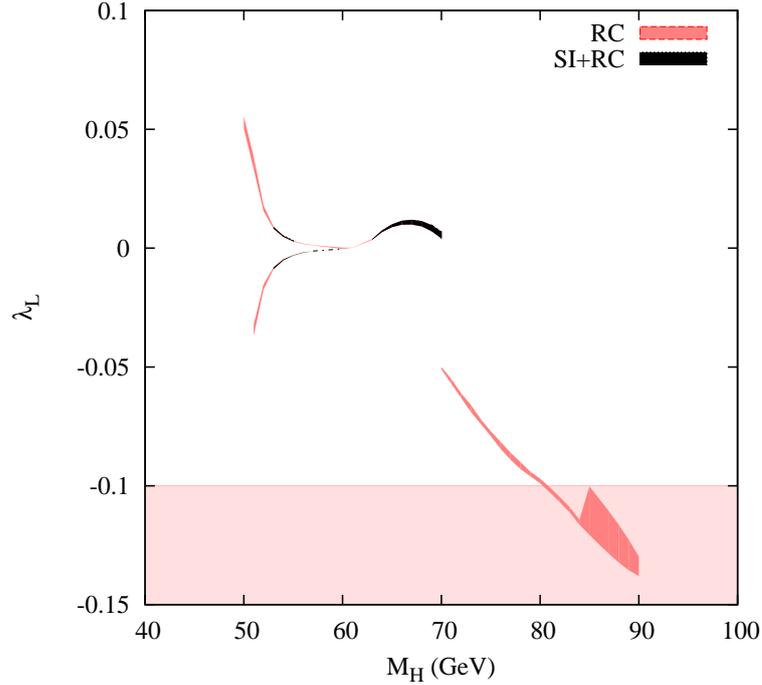}
		\caption{Region allowed by imposing the constraints on relic density(RC) and spin-independent cross-section(SI) for DM-nucleon scattering. 
			The red(gray) region is allowed only by the requirement of $\Omega_{\rm DM}h^2$ being in the correct range. 
			The black region is allowed by both the $\Omega_{\rm DM}h^2$ and direct detection constraints. 
			The shaded horizontal band below is disallowed by vacuum stability conditions. Here, $M_{H^\pm}=M_A=200$ GeV.}
		\label{f:lLMH_DM} 
	\end{figure}
  \item It should be noted
	that in Fig.~\ref{f:lLMH_DM} we keep $M_H^{\pm}$ and $M_A$
	fixed at 200 GeV. For higher values of $M_{H^\pm}$ and $M_A$, there is hardly any change
	in the annihilation cross-section. However, for values of $M_H^{\pm}$ and $M_A$ less than 200 GeV, 
	the allowed region of Fig.~\ref{f:lLMH_DM} gets slightly modified. For example, for $M_H \simeq 70~\rm GeV$
	and $M_{H^\pm}=M_A=200~\rm GeV$, the DM-DM-Higgs coupling $\lambda_L\simeq 0.007$, but for $M_{H^\pm}=M_A=100~\rm GeV$, one needs 
	$\l_L\simeq 0.009$ to satisfy the relic density constraint. It should be noted however that
	both of the above points in the parameter space are within the stability region as shown in Fig.~\ref{f:lLMH}.\\
	Therefore, it is not possible to constrain $M_H^{\pm}$ and $M_A$ using DM constraints alone,
	theoretical constraints however predict strong
	upper bounds on these masses, as is evident from Fig.~\ref{f:l3Mhpm} and Fig.~\ref{f:MC}.
	\begin{figure}[htbp!]
		\begin{center}
			\subfloat[ \label{sf:MHCl3lowM}]{%
				\includegraphics[scale=0.5]{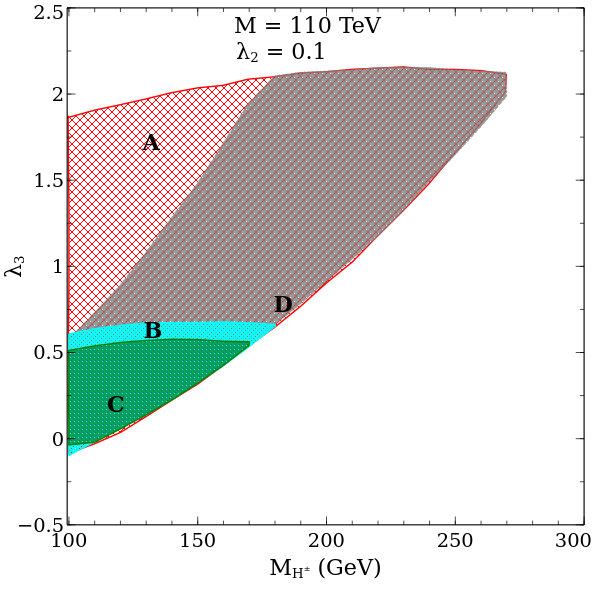}
			}~~~~
			\subfloat[\label{sf:MHCl3lowM}]{%
				\includegraphics[scale=0.5]{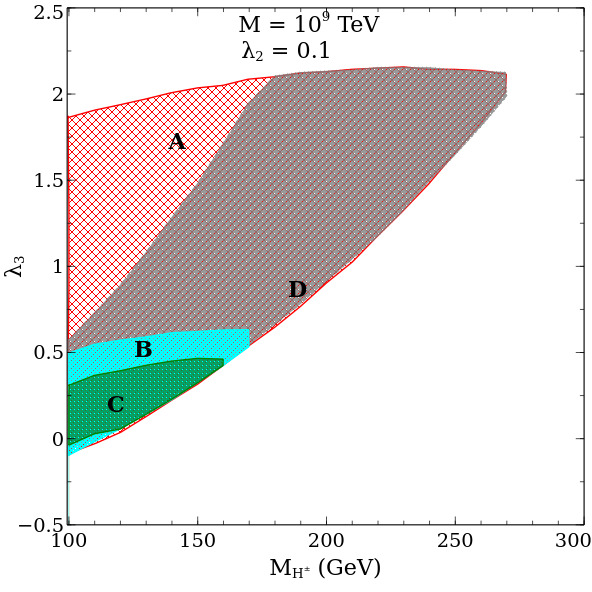}~~~~
			} \\
			\subfloat[ \label{sf:MHCl3lowM}]{%
				\includegraphics[scale=0.5]{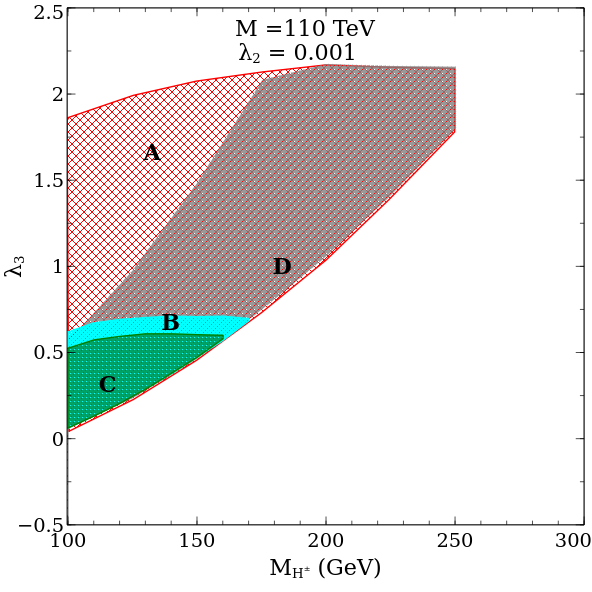}
			}~~~~
			\subfloat[\label{sf:MHCl3lowM}]{%
				\includegraphics[scale=0.5]{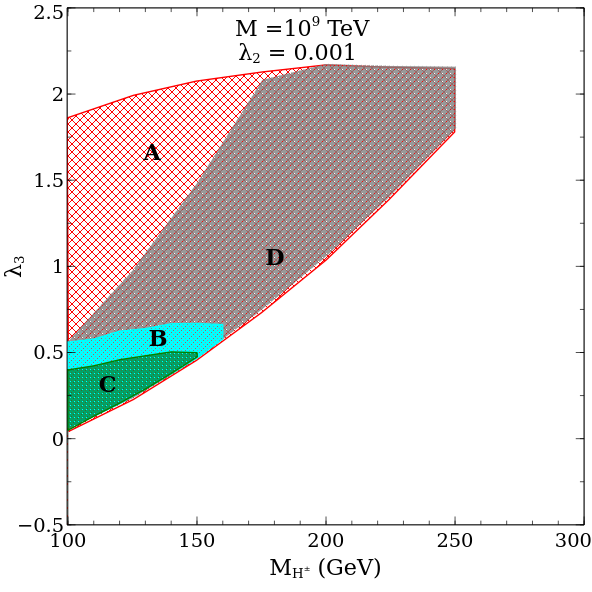}~~~~
			} 			
			\caption{Regions allowed by the theoretical constraints projected in the $M_H^{\pm}$-$\lambda_3$ plane
				for  $M = 110$ TeV (left panel) and $10^{9}$ TeV (right panel) with three different choices 
				of $\Lambda_{UV}$ and two values of $\lambda_2$ (upper and lower panel). 
				The regions denoted by A (red), B (cyan) and 
				C (green) obey those constraints up to $\Lambda_{UV}$ = $10^{6}$, $10^{16}$ and $10^{19}$ GeV, respectively. 
				The grey region denoted by D shows the 2$\sigma$ allowed limit of the Higgs to diphoton signal strength.}
			\label{f:l3Mhpm}
		\end{center}
	\end{figure} 
	\begin{figure}[!htbp]
		\begin{center}
			\subfloat[ \label{sf:MAlLlowM}]{%
				\includegraphics[scale=0.5]{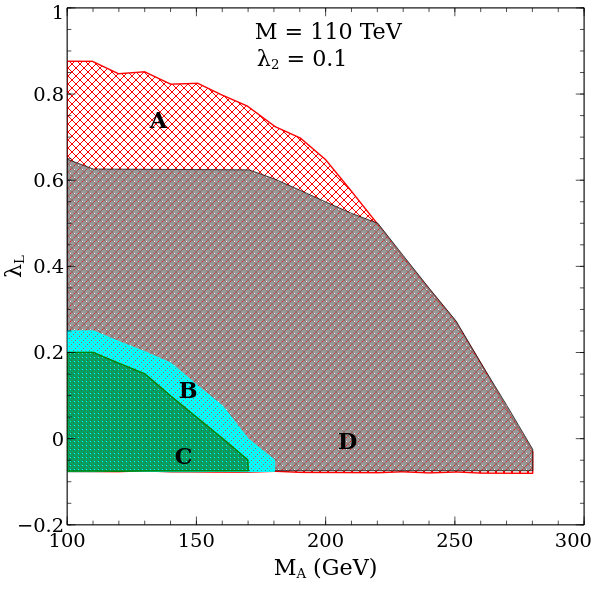}
			}~~~~
			\subfloat[\label{sf:MAlLhighM}]{%
				\includegraphics[scale=0.5]{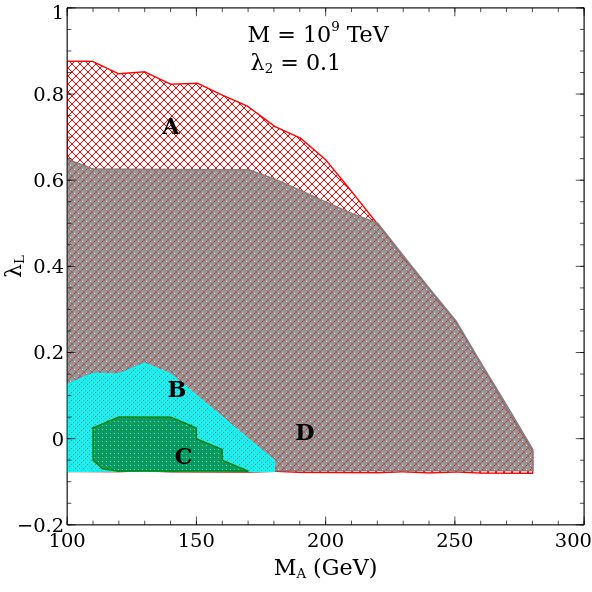}~~~~
			} \\
			\subfloat[ \label{sf:MHCl3lowM}]{%
				\includegraphics[scale=0.5]{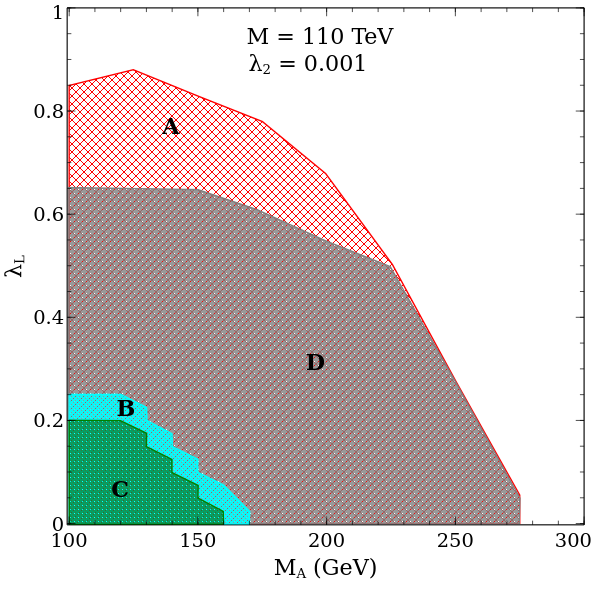}
			}~~~~
			\subfloat[\label{sf:MHCl3lowM}]{%
				\includegraphics[scale=0.5]{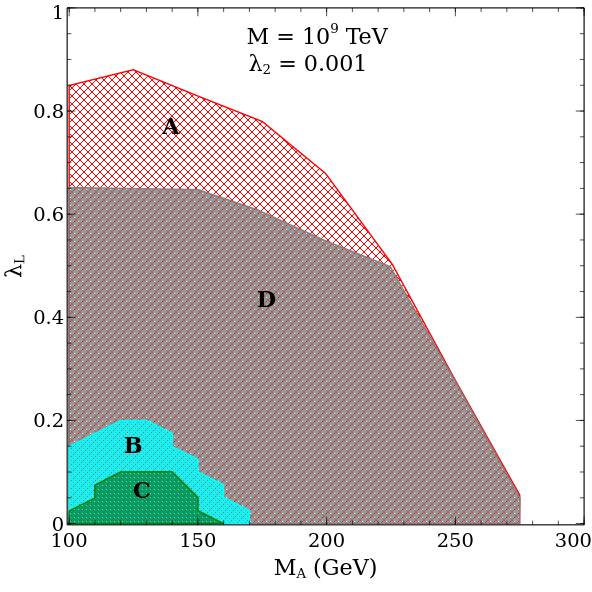}~~~~
			}
			
			\caption{ Regions allowed by the theoretical constraints projected in the $M_A$-$\lambda_L$ plane
				for  $M = 110$ TeV (left panel) and $10^{9}$ TeV (right panel) with three different choices 
				of $\Lambda_{UV}$ and two values of $\lambda_2$ (upper and lower panel). 
				The regions denoted by A (red), B (cyan) and 
				C (green) obey those constraints up to $\Lambda_{UV}$ = $10^{6}$, $10^{16}$ and $10^{19}$ GeV, respectively. 
				The grey region denoted by D shows the 2$\sigma$ allowed limit of the Higgs to diphoton signal strength.}
			\label{f:MC}
		\end{center}
	\end{figure} 
	\item In Fig.~\ref{f:l3Mhpm} and Fig.~\ref{f:MC}, we show the theoretically valid regions in the $\l_3-M_H^{\pm}$ 
	and $\l_L-M_A$ planes. As in previous cases, we exhibit the same for two different $M$ and $\l_2$ values.
	Also, in each case, we overlay the parameter spaces allowed by the $2\sigma$ limit of Higgs to diphoton signal strength (region {\bf D}).  We observe that a tight upper bound of 160-170 GeV is realised on the masses $M_H^{\pm}$ and $M_A$ for the cut-off at the Planck scale and the couplings ($\l_3,\l_L$) are also bounded. 
    The upper bounds on the masses $M_H^{\pm}$ and $M_A$ are imposed by the requirement
	of perturbativity and unitarity up to the desired cut-off. With $M_H$ in the aforementioned range,
	large masses of the other $Z_2$-odd scalars imply high values 
	of the quartic couplings at the electroweak scale which potentially violate perturbativity or unitarity at some high scale. 
    On the other hand, the vacuum stability condition $\l_3 + \sqrt{\l_1 \l_2} \ge 0$ forbids large negative values of $\l_3$ and
    perturbativity puts the upper bound. The explanations of the limits of $\l_L$ have already been mentioned above. 
	\item The new physics contribution to the Higgs to diphoton decay channel comes from the charged Higgs loop which
	is a function of $M_{H^\pm}$ and $\l_3$. The fact that $\l_3$ can not be large negative results in a fall in the signal strength for diphoton channel. However, the parameter space for $\L_{UV} = 10^{19}$ GeV is 
	still allowed by the current limits on $\mu_{\gamma \gamma}$ defined in Eq.~\ref{mu_gamma}. 
	\end{itemize}
\subsection{High mass DM region}
This section demonstrates the high scale validity of our scenario in the limit
of a high DM mass. As discussed earlier, one needs to tune the mass splitting
amongst $H$, $H^\pm$ and $A$ and the coupling $\l_L$ to an appropriate degree
in order to achieve a relic density within the desired bounds. It is seen 
that the maximal allowed splitting $(\Delta M)$ 
amongst the masses of the $Z_2$ odd scalars is 10 GeV.  As previously mentioned,
for each chosen values of $\l_2$ (0.1 and 0.001),
one is thus motivated to scan the high DM mass region in the following 
ranges:
\bea
\l_L: [-1.0,1.0] \\
M_H:  [550.0 ~\rm GeV, 1000.0~\rm GeV] \\
M_A : [ M_H, M_H+10.0 ~\rm GeV]\\
M_{H^+} : [ M_H, M_H+10.0 ~\rm GeV]
\eea

Unlike the previous case, while the theoretical constraints 
ruled out a large portion of the parameter space, the DM constraints put a less stringent bound on it
in this high DM mass region. Therefore, in Fig.~\ref{f:MHlL} and Fig.~\ref{f:MHCl3}, we demonstrate the parameter region allowed
only by the DM constraints and then overlay the part which are further allowed by theoretical constraints.
The full (pink) region denoted by {\bf RC+ SI} shows the valid parameter space 
allowed by DM constraints. 
In accordance with previous notation, the regions denoted by A (red), B (cyan) and 
	C (green) denote $\L_{UV}$ = $10^{6}$, $10^{16}$ and $10^{19}$ GeV respectively.

 Let us explain the various features of the model that emerges from the figures, in detail.   

\begin{enumerate}[$(1)$]
	\item The interplay of the theoretical and experimental constraints is studied in the 
	form of correlation plots in the $M_H$ - $\l_L$ (Fig.~\ref{f:MHlL}) as well as 
	$M_{H^\pm}$ - $\l_3$ (Fig.~\ref{f:MHCl3}) plane. As can be seen, a significant amount of parameter
	space is forbidden for the theory to be valid until the Planck scale.
\item Similar to the previous section, the upper and lower bounds on $\l_L$ are placed from the 
requirements of perturbative unitarity and vacuum stability conditions respectively, as shown in Fig.~\ref{f:MHlL}. In this region too the lower limit of $\l_L$ has dependence on $\l_2$.
 The same happens for $\l_3$, as depicted in Fig.~\ref{f:MHCl3}.
\begin{figure}[!htbp]
	\begin{center}
		\includegraphics[scale=0.5]{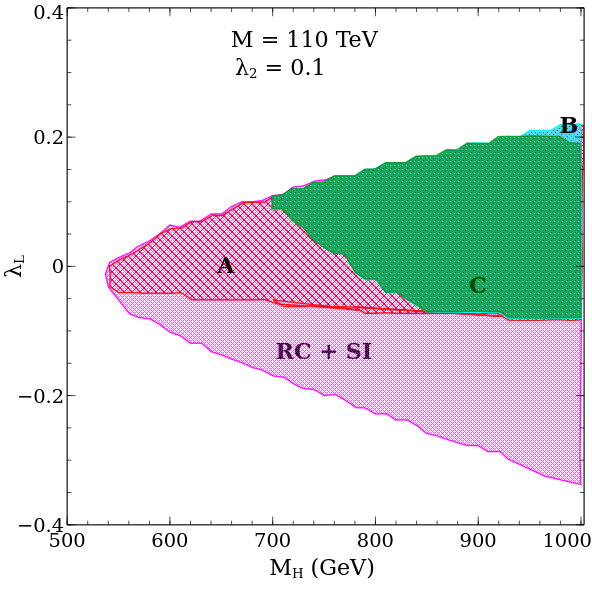}~~~~~~~
		\includegraphics[scale=0.5]{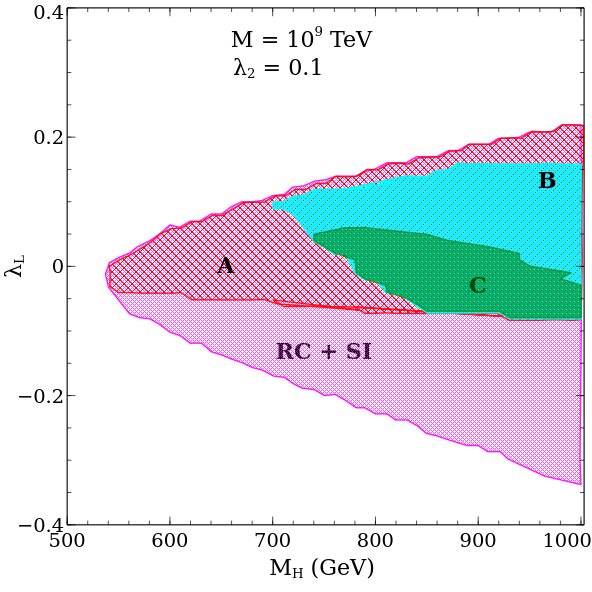}
		\includegraphics[scale=0.5]{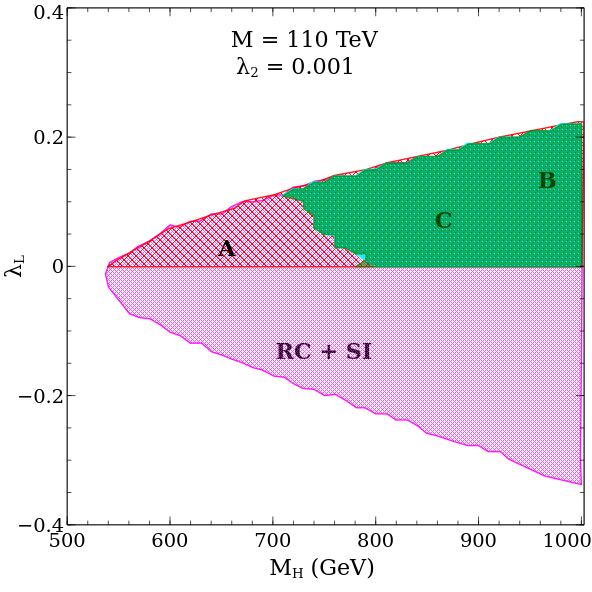}~~~~~~~
		\includegraphics[scale=0.5]{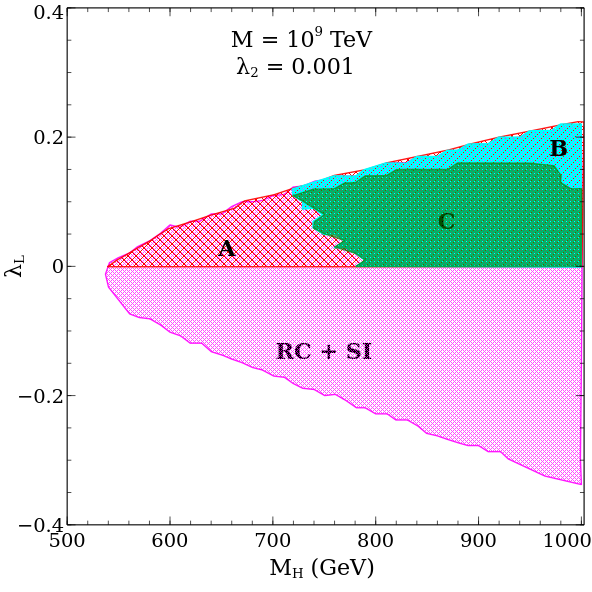}
		\caption{Region(s) allowed in the $M_H$-$\lambda_L$ plane obeying the various constraints
			for $M = 110$ TeV (left panel) and $M = 10^{9}$ TeV (right panel) with three different choices 
			of $\Lambda_{UV}$ and two values of $\lambda_2$ (upper and lower panel). 
			The full region (marked by {\bf RC + SI}) (magenta) is allowed 
			by the DM constraints alone. The overlapped regions labeled by {\bf A} (red), {\bf B} (cyan) and {\bf C} (green)
			are consistent with the theoretical constraints up to $\Lambda_{UV} = 10^{6},~ 10^{16}$ and $10^{19}$ GeV, respectively.}
		\label{f:MHlL}
	\end{center}
\end{figure}
\item It is to be noted
that the theoretically allowed parameter space in this high DM mass
region is fully consistent 
with the Higgs to diphoton LHC data. Since, a heavy $H^{\pm}$ which 
naturally arises in this region, does not 
cause any significant deviation in diphoton signal strength for the SM-like scalar. 
This occurs even with a large positive $\l_3$ (within the bounds
shown in Fig.~\ref{f:MHCl3}). In principle, a large negative $\l_3$ could modify $\mu_{\gamma \gamma}$ unacceptably. 
However, as Fig.~\ref{f:MHCl3} shows, such values are inconsistent with the aforementioned
theoretical constraints.  
\begin{figure}[htbp!]
	\begin{center}
		\includegraphics[scale=0.5]{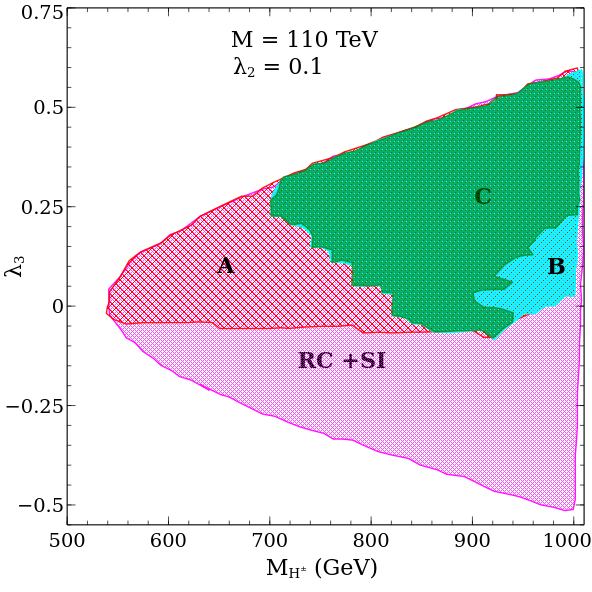}~~~~~~~
		\includegraphics[scale=0.5]{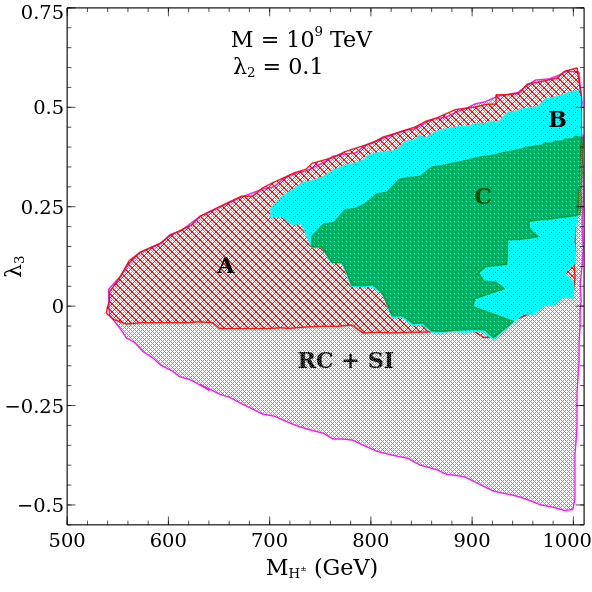}
		\includegraphics[scale=0.5]{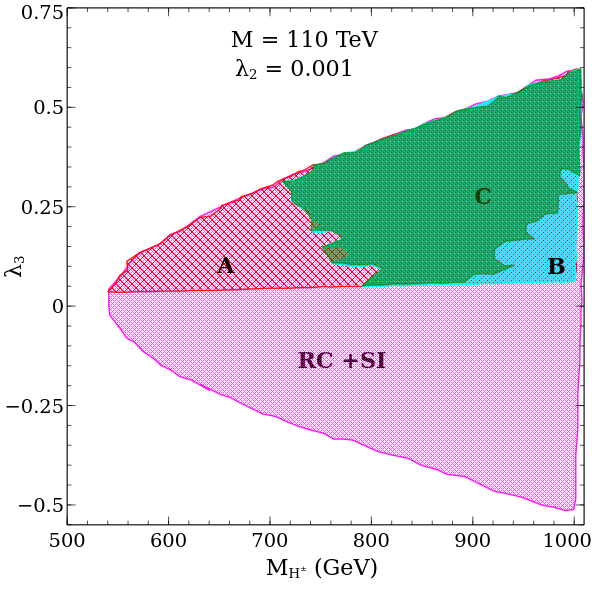}~~~~~~~
		\includegraphics[scale=0.5]{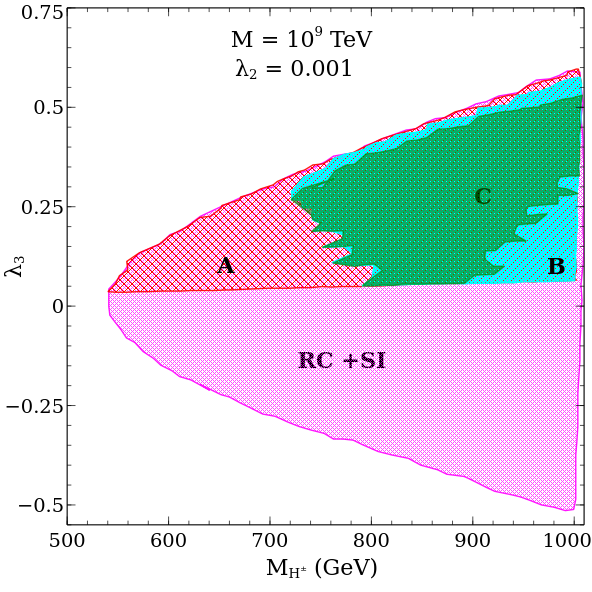}
		\caption{Region(s) allowed in the $M_H^{\pm}$-$\lambda_3$ plane obeying the various constraints
			for $M = 110$ TeV (left panel) and $M = 10^{9}$ TeV (right panel) with three different choices 
			of $\Lambda_{UV}$ and two values of $\lambda_2$ (upper and lower panel). 
			The full region (marked by {\bf RC + SI}) (magenta) is allowed 
			by the DM constraints alone. The overlapped regions labeled by {\bf A} (red), {\bf B} (cyan) and {\bf C} (green)
			are consistent with the theoretical constraints up to $\Lambda_{UV} = 10^{6},~ 10^{16}$ and $10^{19}$ GeV, respectively.}
		\label{f:MHCl3}
	\end{center}
\end{figure}
\item It is worth noting here that the parameter space valid until the Planck scale and
 corresponding to $M=110~ {\rm TeV}$
shrinks significantly for $M=10^{9}~\rm TeV$, as can be seen in Fig.~\ref{f:MHlL} and Fig.~\ref{f:MHCl3}.
 This is because, the significant part of the parameter space is discarded by the stability 
condition at large $M$ where the neutrino Yukawa couplings becomes large ${\cal O}(10^{-1})$. Such 
large Yukawa coupling contributes to the beta function
of $\l_2$ as shown in Eq.~\ref{e:lam2RG} through the terms $+\l_2y_\nu^2$ 
and $-y_\nu^4$ that either makes $\l_2$ perturbative in some cases or $\l_2$ negative and the vacuum unstable
in other.
 We elaborately discuss this feature below with some benchmark points. 
\end{enumerate}

\begin{figure}[!htbp]
\begin{center}
    \subfloat[ \label{sf:RGlowMBP3}]{%
      \includegraphics[scale=0.5]{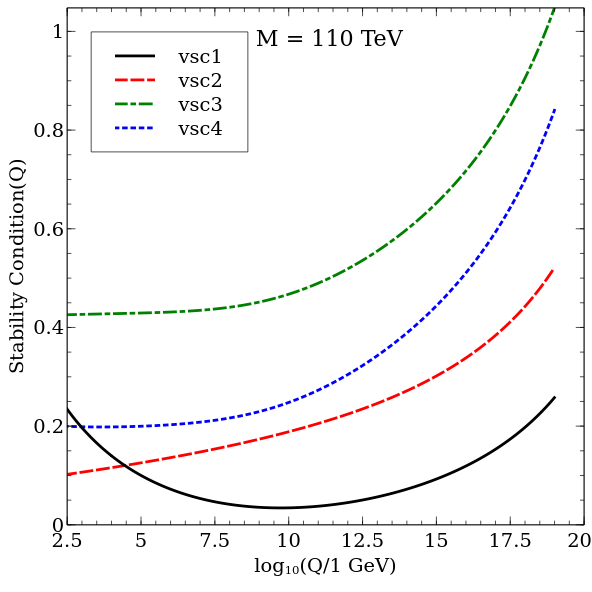}
    }~~~~
    \subfloat[\label{sf:RGhighMBP3}]{%
      \includegraphics[scale=0.5]{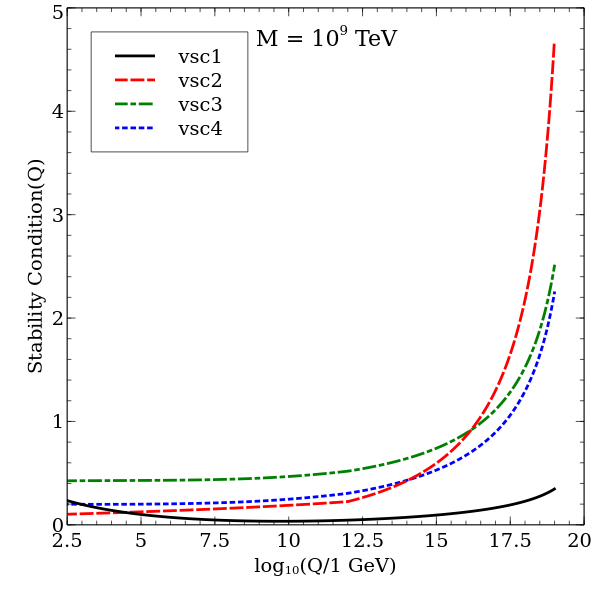}~~~~
    } 
\caption{RG running of different scalar quartic couplings corresponding to BP1. 
The solid, dashed, dashed dotted and dotted lines
denote the evolution curves of the stability conditions
 vsc1, vsc2, vsc3 and vsc4 respectively.}
    \label{f:RGBP3}
\end{center}
\end{figure}

We have selected two benchmark points(BP1 and BP2) as samples out of the allowed
regions consistent with the relic density constraints. These points demonstrate how
the different theoretical constraints can truncate the scale of validity of this
scenario for two different values of right-handed neutrino masses
$M=110$ TeV and $M=10^{9}$ TeV respectively. The parameter values are listed in Table.~\ref{benchmarks}.
\begin{table}[ht!]
\centering
\begin{tabular}{|c|c|c|c|c|c|}
\hline
BP & $M_H$    & $M_{H^\pm}$ & $M_A$   & $\lambda_L$  & $\lambda_2$   \\ \hline \hline
BP1 & 850.0 GeV    & 854.0 GeV     & 858.0 GeV & 0.02 & 0.1 \\ \hline
BP2 & 710.0 GeV     & 712.0 GeV      & 711.0 GeV & 0.11 & 0.1                    
\\ \hline \hline
\end{tabular}
\caption{Benchmark values (BP) of parameters affecting the RG evolution of the quartic couplings.
For each BP, two values of $M$, namely, $110$ TeV and $10^{9}$ TeV, have been used.}
\label{benchmarks}
\end{table}
BP1 and BP2 yield $\Omega_{\rm DM}h^2$ = 0.1151 and 0.1207 respectively, which is within with the
allowed range of relic density.
For $M = 110$
TeV, BP1 ensures a stable vacuum till the Planck scale (Fig.~\ref{sf:RGlowMBP3}). 
It is also consistent throughout with perturbativity and unitarity. However, 
one has $y_{\nu} = 0.168$ at $M=10^{9}$ TeV. For this value,  
the term ${\cal O}(\l_2 y_{\nu}^2)$ has a dominant role in the RG evolution  
and $\l_2$ starts increasing rapidly from $10^{9}$ TeV on wards (Fig.~\ref{sf:RGhighMBP3}).  

\begin{figure}[!htbp]
\begin{center}
    \subfloat[ \label{sf:RGlowMBP4}]{%
      \includegraphics[scale=0.5]{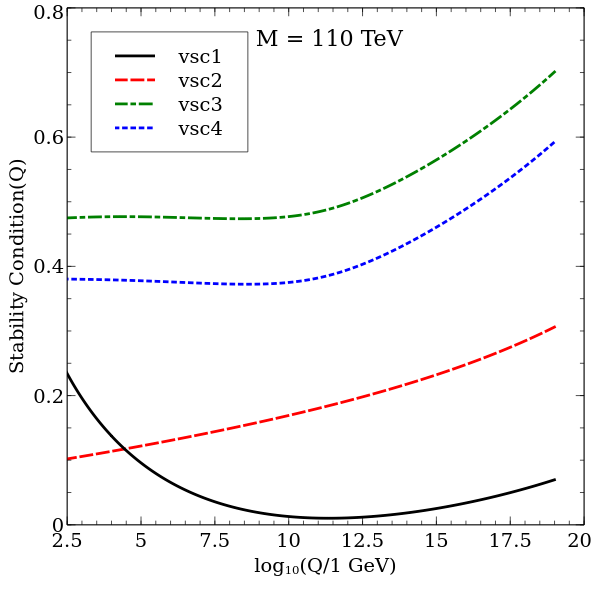}
    }~~~~
    \subfloat[\label{sf:RGhighMBP4}]{%
      \includegraphics[scale=0.5]{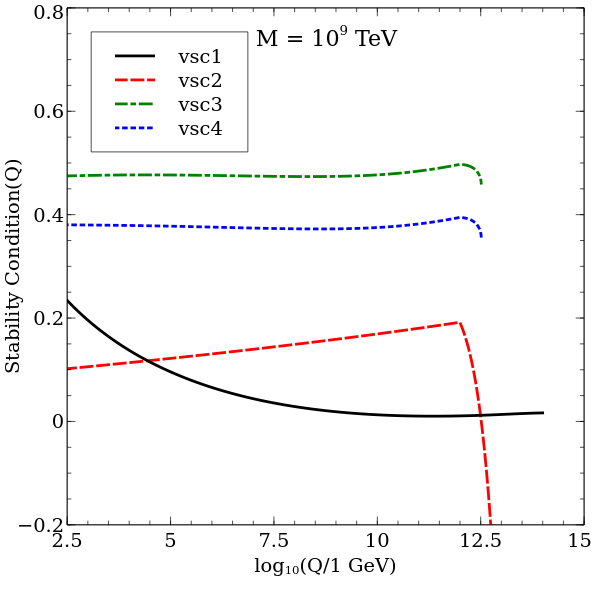}~~~~
    } 
\caption{Same as Fig.~\ref{f:RGBP3} but for the benchmark point BP2.}
    \label{f:RGBP4}
\end{center}
\end{figure}

For $M=110$ TeV, BP2 exhibits similar features in the evolution trajectory as in BP1 (Fig.~\ref{sf:RGlowMBP4}).
For $M = 10^{9}$ TeV, the Yukawa coupling $y_{\nu}$ starts with an initial value around
0.51. This is accounted for by the very small mass splitting, of the order of a GeV, 
between $H$ and $A$. Thus, the dominant contribution from the RH neutrinos comes from
the ${\cal O}(y_{\nu}^{4})$ term that causes $\l_2$ to drop sharply (Fig.~\ref{sf:RGhighMBP4}). Hence, in BP2,
vacuum stability is destroyed shortly after $10^{9}$ TeV as the condition $\l_2 > 0$ gets violated. 
This particular feature can only be witnessed in the case of 
closely spaced $M_H$ and $M_A$, which is the primary requirement to satisfy the relic density constraints
discussed before. This completes the explanation of how the allowed area can shrink
due to different theoretical constraints for $M=10^{9}$ TeV. 
\section{Summary and Conclusions}\label{conclusions}
We have examined the high-scale validity of a scenario that 
(a) offers a scalar dark matter, (b) radiatively generates Majorana
masses for neutrinos, and (c) is responsible for leptogenesis.  
For this, we extend the SM fields with
one additional inert Higgs doublet field $(\Phi_2)$ 
and three right handed neutrinos $(N_i)$. These new particles are odd under
a discrete $Z_2$ symmetry, while all the SM particles are even. Because 
of this discrete symmetry, $\Phi_2$ does not acquire any vacuum expectation 
value (vev) and has no tree-level couplings to fermions. In this scenario,
one has five physical scalars $(h, H,A, H^\pm)$, where, $h$ is denoted 
as the SM like Higgs boson with a mass of 125 GeV. The lightest state between
$H$ and $A$ is the dark matter candidate due to built in $Z_2$ symmetry.
In our analysis we have assumed $H$ to be the dark matter candidate.
Neutrino masses are generated at the one-loop level. The neutrino masses
and mixing angles are determined in terms of Yukawa couplings 
$(y_{\nu})$, new Higgs particle masses $(M_H, M_A)$ and three heavy 
Majorana masses $(M_{1,2,3})$. In our numerical analysis we have assumed
$M_1$ is mass of the lightest state and considered two values, namely, 
$M_1\equiv M = 110 $ TeV and $10^{9}$ TeV. These two mass scales are
consistent with Leptogenesis . For simplicity, in our analysis, we have 
considered only one diagonal Yukawa coupling and to determine the value of this
coupling we have scanned over $M_H$ and $M_A$ for a given value of $M$, 
by keeping $M_\nu \sim {\cal O}(0.1~{\rm eV})$. 

The parameter space of this model is determined in terms of 
additional Higgs boson masses, $M_H, M_A, M_{H^\pm}$, one 
quartic coupling $\lambda_2 $ and a set of 
quartic coupling combinations, $\lambda_L$. While scanning the
parameter space of this model, we have imposed the LEP bound on scalar 
masses, $M_H,M_A$ and $M_{H^\pm}$. 

As far as the dark matter constraints are concerned,
we have used the relic density limits obtained at $3\sigma $ uncertainty 
from the PLANCK experiment and the direct detection cross-section limit
from the LUX experiment. Finally,
for $M_H < M_h/2$, which would lead to large invisible decay width of 
the SM like Higgs boson, we demanded that the corresponding branching 
ratio is less than $20\%$ as obtained from the model independent 
Higgs precision analysis. 

With these set of constraints in hand, we have then scanned the IDM parameter space
for two different ranges of dark matter masses, 
$50~{\rm GeV} < M_H < 90~{\rm GeV}$ and $M_H > 500$ GeV. It should be
noted that with $M_h$ fixed at 125 GeV, $\lambda_2, \lambda_L, M_H, M_A $
and $M_{H^\pm}$ determined all the remaining quartic couplings. We have used
these quartic couplings as the electroweak boundary condition by setting 
the starting RG running scale $Q = M_t$ and evolved these couplings up to 
the scale $\Lambda_{\rm UV}$, where this scenario remained consistent. 
In the RG evolution of these quartic couplings, the neutrino Yukawa couplings 
started playing its role from the right handed neutrino mass scale $Q = M$ 
on wards.  Following are the salient features of this model
	that our analysis brings out.

\begin{itemize}
	\item In our study we have explicitly demonstrated
that at the low DM mass region, the vacuum stability, 
perturbativity and unitarity constraints put stringent limits
on the low-energy value of the coupling $\l_L$ and the $Z_2$-odd scalar masses $M_{H^\pm}, M_A$.
These bounds strongly depend upon the scale up to which the theory is 
valid and the right-handed neutrino mass scale. It is interesting 
to note that all the parameter space allowed by the DM relic density and 
direct detection constraints lies well within the region allowed by the 
theoretical constraints valid up to the Planck scale. However, 
once we have imposed the condition that the Higgs to diphoton signal strength $(\mu_{\gamma \gamma })$ 
should lie within $2\sigma $ of the weighted average value 
of the ATLAS and CMS data,
the allowed parameter space further squeezed. 
\item The scenario with high DM mass region $(M_H > 500~\rm GeV)$ is 
significantly different from that of the low DM mass region. In this case,
the DM being very heavy, the constraints from direct detection is rather 
insignificant. On the other hand, the relic density constraint is ensured 
by a degenerate $Z_2$-odd scalars $(\Delta M \simeq 10~\rm GeV)$. 
As a result of these, a large part of the parameter space in 
$\lambda_L - M_H $ and $\lambda_3 - M_{H^\pm}$ planes remain unconstrained.
\item However, the study of the high scale validity of high DM mass region
has interesting consequences. The DM-allowed region reduces
substantially after imposing the theoretical constraints and this reduction
is strongly dependent on the cut-off scale $(\Lambda_{\rm UV})$. The effect of neutrino Yukawa couplings in 
the RG evolution of the different quartic couplings are also prominent in this 
case. We have found that with the increase in the right handed neutrino mass 
scale $M$, the neutrino Yukawa coupling $(y_\nu)$ also increases, 
which in turn further reduces the allowed parameter space by either 
destabilizing the vacuum or violating the perturbativity 
bound. There is nonetheless a clearly identifiable region of the parameter space,
which keeps the model valid all the way up to the Planck scale. This scenario  
is consistent with the measured Higgs-to-diphoton rates as measured during the
8 TeV run of the LHC.
\end{itemize}
\section*{Acknowledgments\,:}
The work of NC and BM was partially supported
by funding available from the Department of Atomic Energy, 
Government of India, for the Regional Centre for Accelerator-based Particle Physics (RECAPP), 
Harish-Chandra Research Institute. 
DKG and IS acknowledge the hospitality of RECAPP and High Energy Physics Division, ICTP
during the early stage of
this project, while NC and BM thank the Department of Theoretical Physics, Indian
Association for the Cultivation of Science at a later stage.

\section*{Appendix}
\appendix
\label{s:appen}
\section{One-loop Renormalization group (RG) equations}
\label{ss:RGE}
The RG equations for the gauge couplings, for this model, are given by \cite{Branco:2011iw},
\besub
\bea
16\pi^2 \frac{dg_s}{dt} &=& - 7 g_s^3,
\\
16\pi^2 \frac{dg}{dt} &=& - 3 g^3,
\\
16\pi^2 \frac{dg^{\prime}}{dt} &=& 7 {g^\prime}^3.
\eea
\eesub
Here $g^{\prime}$, $g$ and $g_s$ denote the U$(1)$, SU$(2)_L$ and SU$(3)_c$ gauge couplings respectively.

The quartic couplings $ \lambda_{i} ~(i=1,\ldots,5)$ evolve according to, 
\besub
\bea
\label{e:lam1RG}
16\pi^2 \frac{d\lambda_1}{dt} &=&
12 \lambda_1^2 + 4 \lambda_3^2 + 4 \lambda_3 \lambda_4 + 2 \lambda_4^2
+ 2  \lambda_5^2 + \frac{3}{4}(3g^4 + g^{\prime 4} +2 g^2 g^{\prime 2})\nonumber \\
 & & -\lambda_1 (9 g^2 + 3 g^{\prime
2} - 12 y_t^2 - 12 y_b^2 - 4 y_{\tau}^2 ) - 12 y_t^4\,, \\
\label{e:lam2RG}
16\pi^2 \frac{d\lambda_2}{dt} &=&
12 \lambda_2^2 + 4 \lambda_3^2 + 4 \lambda_3 \lambda_4 + 2 \lambda_4^2
+ 2 \lambda_5^2  \nonumber \\
 & &
+\
\frac{3}{4}(3g^4 + g^{\prime 4} +2g^2 g^{\prime 2}) -3\lambda_2
(3g^2 +g^{\prime 2}- \frac{4}{3} y_{\nu}^2)- 4 y_{\nu}^4\,, \\
16\pi^2 \frac{d\lambda_3}{dt}  &=&
\left( \lambda_1 + \lambda_2 \right) \left( 6 \lambda_3 + 2 \lambda_4 \right)
+ 4 \lambda_3^2 + 2 \lambda_4^2
+ 2 \lambda_5^2
+\frac{3}{4}(3g^4 + g^{\prime 4} -2g^2 g^{\prime 2}) \nonumber \\
& & - \lambda_3
(9g^2 + 3g^{\prime 2}- 6 y_t^2 - 6 y_b^2 - 2 y_{\tau}^2- 2 y_{\nu}^2)\,, \\
16\pi^2 \frac{d\lambda_4}{dt}  &=&
2 \left( \lambda_1 + \lambda_2 \right) \lambda_4
+ 8 \lambda_3 \lambda_4 + 4 \lambda_4^2
+ 8 \lambda_5^2 
+\ 3g^2 g^{\prime 2} \nonumber \\
& &- \lambda_4 (9g^2 + 3g^{\prime
2}- 6 y_t^2 - 6 y_b^2 - 2 y_{\tau}^2 - 2 y_{\nu}^2)\,,
\label{eq:dl4} \\
16\pi^2 \frac{d\lambda_5}{dt}  &=&
\left( 2 \lambda_1 + 2 \lambda_2 
+ 
8 \lambda_3 + 12 \lambda_4 \right) \lambda_5
- \ \lambda_5 (9g^2 + 3g^{\prime 2} - 6 y_b^2 - 2 y_{\tau}^2- 6 y_t^2- 2 y_{\nu}^2)\,, \label{eq:dl5}  
\eea
\eesub
For the Yukawa couplings the corresponding set of RG equations are,
\besub
\bea
 16\pi^2 \frac{dy_b}{dt} &=& y_{b}\left(-8g_s^2 - \frac94 g^2 - \frac{5}{12} g^{\prime 2}+
 \frac92 y_{b}^2 +y_{\tau}^2 + \frac32 y_{t}^2\right)\,,\\
 16\pi^2 \frac{dy_t}{dt} &=& y_{t}\left(-8g_s^2 - \frac94 g^2 - \frac{17}{12} g^{\prime 2}+
 \frac92 y_{t}^2 +y_{\tau}^2 + \frac32 y_{b}^2\right)\,,\\
 16\pi^2 \frac{dy_{\tau}}{dt} &=& y_{\tau}\left(-\frac94 g^2 - \frac{15}{4} g^{\prime 2}+ 3y_{b}^2 + 3y_{t}^2 + \frac12 y_{\nu}^2 + \frac52 y_{\tau}^2\right )\,.\\
  16\pi^2 \frac{dy_{\nu}}{dt} &=& y_{\tau}\left(-\frac94 g^2 - \frac{3}{4} g^{\prime 2} - \frac34 y_{\tau}^2 + \frac52 y_{\nu}^2\right)\,.
\eea
\eesub


\bibliographystyle{JHEP}
\bibliography{ref.bib}       
\end{document}